# Photonic crystal nanobeam lasers

Y. Zhang,[1,a)] M. Khan,[1] Y. Huang,[2] J. Ryou,[2] P. Deotare,[1] R. Dupuis,[2] and M. Lončar[1]

[1] School of Engineering and Applied Sciences, Harvard University
[2] School of Electrical and Computer Engineering, Georgia Institute of Technology
a) Electronic mail: yinan@seas.harvard.edu

**Abstract**: Photonic crystal lasers operating at room temperature based on high $Q/V$ nanobeam cavities have been demonstrated. We reported a large spontaneous emission factor ($\beta \sim 0.97$) by fitting the *L-L* curve with the rate equations.

Photonic crystal lasers,[1] with small mode volumes ($V$) and high Quality ($Q$) factors, have enhanced photon emission below threshold via Purcell effect,[2-4] and can operate with high modulation speeds.[5,6] Moreover, threshold-less lasing has been predicted in photonic crystal cavities, with necessary but not sufficient condition that the spontaneous emission factor ($\beta$) amounts to unity.[7,8] Most photonic crystal lasers so far have been demonstrated based on two-dimensional (2D) photonic crystal slabs. Recently nanobeam structures attracted extensive interests because of their capability to achieve ultra-high $Q/V$ factors in much smaller footprint.[9-11] We emphasize a small mode volume of these resonators, close to the diffraction limit [$(\lambda/2n)^3$], that is crucial for realization of large Purcell factor in solid-state emitters with a considerable homogeneous broadening (e.g. bulk semiconductors, semiconductor quantum wells at room temperature). In addition, nanobeam cavities do not have mode degeneracy,[12-15] and therefore can support a single cavity mode over a broad spectrum. This single-mode nature is important for a large $\beta$ factor and reduction of lasing threshold[8]. In this work we report the experimental demonstration of photonic crystal nanobeam lasers operating at room temperature.

The nanobeam cavities are designed using the same approach as our previous work.[9] The cavity is designed, using three dimensional finite-difference time-domain (3D-FDTD) modeling, to support a fundamental TE-like mode at 1.59μm, polarized predominantly along *x*-axis. Field components of the fundamental mode are shown in Fig. 1(a). The electric field density is concentrated in the dielectric region, leading to a large confinement factor needed for a lasing action. The mode volume of this mode is $0.28(\lambda/n)^3$, which is close to the diffraction limit, and its passive $Q$ factor is larger than 8,000,000. In addition, the cavity supports another mode at a longer wavelength (1.72μm). This is an extended mode, with a node in the center of the structure and a larger mode volume of $0.67(\lambda/n)^3$.

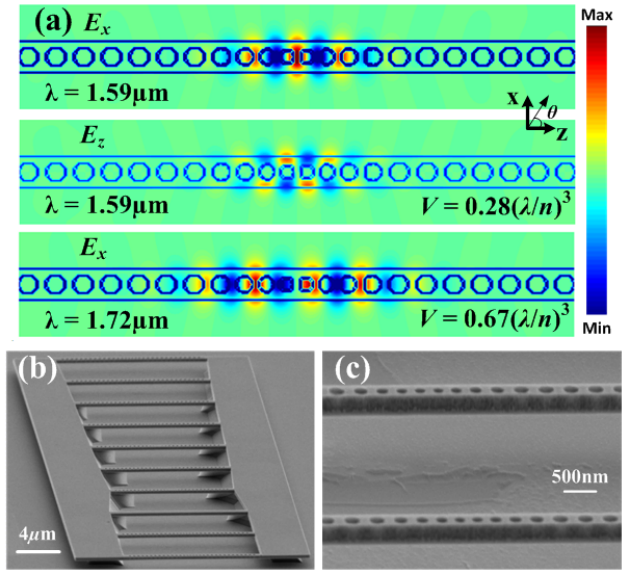

FIG. 1 (a) The lasing mode profile ($E_x$ and $E_z$ components) that resonates at 1.59μm. The higher-order mode profile ($E_x$ component) that resonates at a higher wavelength of 1.72μm. (b)(c) Scanning electron micrograph image (side-view) of the final fabricated nanobeam structures with low and high magnification.

Our cavities are fabricated in a 330nm thick, MOCVD grown, $In_{0.53}(Al_{0.4}Ga_{0.6})_{0.47}As$ slab that contains four compressively strained $In_{0.58}Ga_{0.42}As$ quantum wells placed at the center of the slab. Active layer's emission is maximized at wavelength of 1.59μm, and is transverse-electric (TE) polarized due to the strain. The slab is grown on top of a 1μm thick sacrificial InP layer. Top-down fabrication sequence, based on e-beam lithography (using negative Foxx resist) followed by ICP reactive ion etching (BCl$_3$/HBr chemistry) is used to realize the structures. The remaining mask layer is then removed in HF, and nanobeams are released in 3:1 HCl:H$_2$O solution at 11ºC. The final fabricated structures are shown in Fig. 1(b). The pattern consists of 10 nanobeam supported by the two pads located aside.

The samples are photo-pumped at room temperature with a 660nm pulsed laser diode. The pulse width and repetition rate are 9ns and 300kHz respectively, corresponding to a duty cycle of 0.27%. The pump beam is focused to a ~4μm diameter spot via a 100X objective lens, and the emitted light is collected via the same objective lens and analyzed using optical spectrum analyzer (OSA), infrared camera, and IR detector.

In Fig. 2(a), we show the output power as a function of the peak pump power (*L-L* curve) incident onto the nanobeam for one of the cavities. Soft turn-on of the laser, without pronounced threshold, is indication of a large $\beta$ factor. In order to unambiguously attribute the lasing action to the localized defect mode, we used NIR camera to image the mode profiles at different pump levels (Fig. 2(a)). It can be seen that the emission spot is well localized to the center of the beam. In addition, by scanning the sample in x-z plane using a piezo-actuated stage with a spatial resolution of 6nm, we checked the dependence of the output power on the pump beam position [Fig. 2(b)]. As the pump beam is moved away from the center of the cavity, the beam intensity decreases rapidly and finally vanishes. This is further confirmation of emission from localized mode of the cavity, and not extended band-edge emission. Here, the slightly elongated shape along x-axis of the emission profile is due to the elliptical shape of the pump beam used in the experiment: the waist of pump beam is larger in x-direction, and it is much larger than the width of the nanobeam. This also allows us to evaluate the effective pump power, that is, the overlap of the pump beam with the nanobeam. This effective pump power is reported in Fig. 2(a). We evaluate an effective threshold of 84$\mu$W for our nanobeam laser. We note, however, that the threshold is even smaller, since not all the pump power incident onto the nanobeam is absorbed by the cavity. Finally, in Fig. 2(c) we show the polarization dependence of the emission beam. The laser is polarized in x-axis and exhibits a large polarization ratio of ~10, which is another signature of the highly polarized cavity mode.

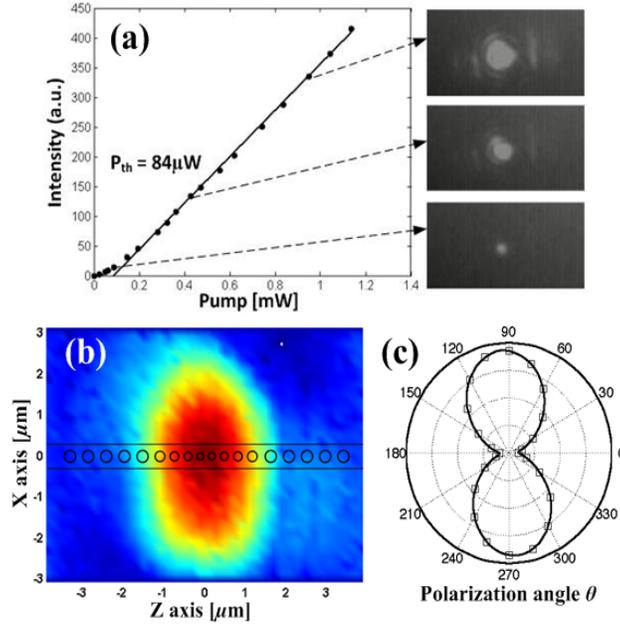

FIG. 2 (a) Laser emitted power as a function of the incident power (*L-L* curve). The spectrum of the mode at transparent point is shown in the inset. The emission profiles obtained from the camera at different pump levels are shown on the right. (b) Output lasing power as a function of the pump beam position. The pattern of the nanobeam is superimposed as the background of the picture (c) Polarization dependence of the lasing mode. The angle $\theta$ is defined in Fig. 1(a).

The carrier and photon behaviors of the photonic crystal lasers can be analyzed using the following rate equations[16],

$$\begin{cases} \dfrac{dN}{dt} = \tilde{F}_{in} - \Gamma G(N)P - AN - (F_{sp,0} + \alpha)BN^2 - CN^3 \\ \dfrac{dP}{dt} = \Gamma G(N)P - \dfrac{P}{\tau_c} + F_{sp,0}BN^2 \end{cases} \quad (1)$$

where $N$ and $P$ are the carrier and photon density respectively, $\Gamma$ is the confinement factor of the lasing mode, $G(N)$ is the logarithmic gain, and $\tilde{F}_{in}$ is the pumped photon density flux (unit: #/$m^3$/s). The output lasing photon density flux is expressed as $\tilde{F}_{out} = P/\tau_c$, where the photon cavity lifetime $\tau_c$ is related with the $Q$ factor of the lasing mode $\tau_c = Q\lambda_c/c$. In the first equation, $AN$, $(F_{sp,0}+\alpha)BN^2$, and $CN^3$ represent the carrier loss rate through surface recombination, radiative recombination and Auger recombination, respectively. Within the radiative recombination, the emission to the lasing cavity mode is proportional to its Purcell factor $F_{sp,0}$, and the emission to all the non-lasing channels is represented by a coupling factor α. The spontaneous emission factor $β$ can thus be expressed as $\beta = \dfrac{F_{sp,0}}{F_{sp,0}+\alpha}$.[17] Here, α is segregated into two contributions $\alpha = \zeta + \sum_{i\neq 0} F_{sp,i}$. The first contribution includes coupling to the leaky modes that are not confined to the cavity. For a large cavity, the emitter's emission rate to radiative channels is close to its emission in a homogeneous medium ($\zeta \approx 1$). The second factor accounts for coupling to all (non-lasing) cavity modes that have a spectral overlap with the emitter. This coupling, either due to splitting of mode degeneracy[13-15] or large mode volume,[18, 19] has been the origin of low $β$ factor. This second contribution can be greatly suppressed in small mode volume photonic crystal nanobeam cavity that has a free spectral range (spacing between the lasing mode and the closest cavity mode) much larger than the half-linewidth of the electron transition.

For our nanobeam cavities, $\zeta$ is evaluated as 0.91 through FDTD simulation. The contribution to the higher-order cavity mode shown in Fig. 1(a) can be neglected, since this higher-order mode lies ~130nm away from the lasing mode, which is much larger than half of the homogeneous broadening of the quantum well (~4nm).[3] We evaluated the lasing mode's Purcell factor of 26 through the equation $F_{sp} = \eta_{spat}\eta_{pol}\dfrac{3}{4\pi^2}\dfrac{\lambda}{\Delta\lambda_M}\dfrac{1}{V/(\lambda/n)^3}$, where $\Delta\lambda_M \approx \max\{\Delta\lambda_e, \Delta\lambda_c\}$ if one of the broadenings from the electron transition and cavity resonance is much larger than the other, and $\eta_{spat}$ and $\eta_{pol}$ take into account the spatial and polarization overlap between the active medium and the optical mode respectively.[17] From the above calculations, we find that the nanobeam cavity supports a high $β$ factor of 0.97.

Next, we compare the measured $L$-$L$ curve with the curve deduced from the rate equations. Typical values of parameters for III-V material-based quantum wells at room temperature are employed.[16] The cavity $Q$ factor is measured from the full-width half-maximum (FWHM = 0.11nm) at near threshold pump level, as shown in Fig. 3(a). This corresponds to a $Q$ factor of 15,000, and is likely limited by the resolution of our OSA (0.1nm). We expect that actual $Q$ is larger than this value. Fig. 3(b) shows the measured $L$-$L$ in log-log scale along with curves obtained from rate equations for different Purcell factors ($F_{sp}$) and spontaneous emission factors ($β$). It can be seen that the experimental data are in excellent agreement with our theoretical prediction of $F_{sp}$~25 and $β$=0.97.

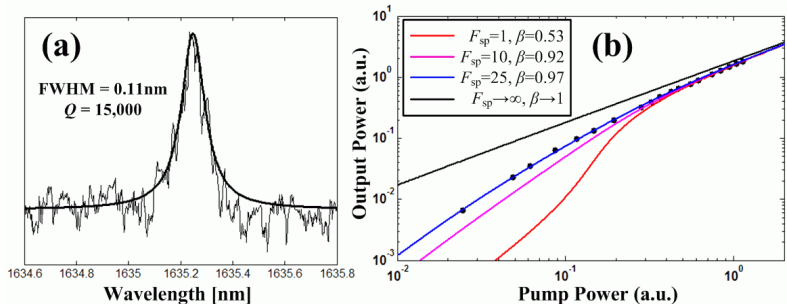

FIG. 3 (a) Laser mode spectrum measured from OSA with narrow resolution (0.1nm) at near threshold pump level. (b) Log-log plot of the L-L curve, with predictions from the rate equations using different $β$ values.

We also show in Fig. 3 the theoretical $L$-$L$ curve as Purcell factor approaches infinity. In this scenario, there is no clear kink as the lasing action is built up. However, we point out that the lasing threshold in this limit is still finite, and is determined by α. In fact, we note that the threshold power is independent of Purcell factor. The threshold photon density flux can be expressed as $\tilde{F}_{in,th} = AN_s + \alpha BN_s^2 + CN_s^3$, where $N_s$ satisfies $G(N_s) = \frac{c}{\lambda_c \Gamma Q}$. Some previous works have drawn similar conclusions.[20, 21] To reduce the lasing threshold or moreover achieve threshold-less behavior, it is necessary to engineer the cavity with small α factor (and thereby a high *β* factor) and low surface recombination. This will be direction taken in our future work.

In conclusion, we have experimentally demonstrated photonic crystal nanobeam lasers operation at room temperature. The small cavity mode volume results in large Purcell factor. This effect, alongside with the single-mode coupling within the electron transition spectrum, leads to a high spontaneous emission factor ~0.97. These results present a promising candidate for low-threshold high-speed nanoscale lasers.

**Acknowledgments**
This work was supported in part by NSF grant ECCS – 0708905 "NIRT: Photon and Plasmon Engineering in Active Optical Devices based on Synthesized Nanostructures", and NSEC at Harvard.